\documentclass[twocolumn,aps,prb]{revtex4} 
\usepackage{graphics,amsmath,graphicx} 
\usepackage{psfig}

\begin{document} 
\title{Surface instability and isotopic impurities in quantum solids}

\author{E. Cappelluti$^{1,2}$, G. Rastelli$^{2,3}$, S. Gaudio$^2$,
and L. Pietronero$^{1,2}$} 

\affiliation{$^1$SMC Research Center and ISC, INFM-CNR, v. dei Taurini 19,
00185 Rome, Italy,}

\affiliation{$^2$Dipartimento di Fisica, Universit\`a ``La Sapienza'',
P.le A. Moro 2, 00185 Rome, Italy}

\affiliation{$^{3}$ Laboratoire de Physique et Mod{\'e}lisation 
des Milieux Condens{\'e}s,   Universit{\'e} Joseph Fourier \\
CNRS - UMR 5493, BP 166, 38042 Grenoble, France}

\begin{abstract} 
In this paper we employ a self-consistent harmonic approximation
to investigate surface melting and local
melting close to quantum impurities in quantum solids.
We show that surface melting can occur at temperatures much lower
than the critical temperature $T_c$ of the solid phase instability in
the bulk. Similar effects are driven by the presence
of an isotope substitution. In this latter case, we show that
stronger local lattice fluctuations,
induced by a lighter isotope atom, can induce local melting
of the host bulk phase.
Experimental consequences and the possible relevance in
solid helium are discussed.
\end{abstract} 
%\pacs{67.80.Cx, 68.08.Bc, 63.20.Ry}
\date{\today} 
\maketitle 

\section{Introduction}

Although melting is a very common phenomenon in nature,
the debate about its microscopic mechanism
is still open.\cite{dash,lowen,forsblom}
The first empirical theory was advanced by Lindemann.\cite{lindemann}
According to this view, melting occurs when the ratio between
the root mean square (rms) $u_{\rm rms}=\sqrt{\langle u^2\rangle}$
of the thermally activated lattice fluctuations
and the lattice constant $a$
exceeds a phenomenological threshold $u_{\rm rms}/a \gtrsim 0.16$
which is roughly material independent \cite{gilvarry,ross}.
In spite of its several flaws (melting is described in terms of the
properties of only the solid phase; no cooperative process
and no role of defects are considered\ldots),
this simple criterion seems to work reasonably well for a variety
of materials.\cite{grimvall_ps}
Lindemann criterion has been recently found to apply as well
at a {\em local} level around crystal defects.\cite{jin,yodh}
This large range of validity of the Lindemann criterion
suggests thus that a microscopic mechanism is actually operative.

The most simple (and employed) model to account for the
Lindemann phenomenology is the self-consistent harmonic approximation (SCHA).
This maps an anharmonic phonon model in a harmonic one.
Anharmonicity is, then, taken into account, at a mean-field level,
through a Debye-Waller-like term which is evaluated self-consistently.
The breakdown of this approach is interpreted as a signal
of solid phase instability, and hence related to  melting.
One of the strength of this theory is that it predicts,
in contrast with the Born criterion but
in agreement with the experimental observation, a partial
but not total softening of the elastic constants of the bulk.

The SCHA represents moreover an efficient tool to understand
in a qualitative way the phenomenon of the surface melting (SM), 
as first proposed by Pietronero and Tosatti (PT).\cite{pt}
In this context, the physical mechanism underlying the surface melting
is quite simple: atoms close to the surface have larger lattice
fluctuations due to the reduced number of nearest neighbor sites, and the
SCHA breaks down consequently at smaller temperatures than in the bulk.
It is clear that this simple theory does not represent an exhaustive
description of the surface melting phenomenology, which should include
roughening, preroughening, partial wetting, the role of ``crystallinity''
etc.\cite{trayanov,tosatti-review} In addition it should be stressed that
the SCHA does not determine directly the melting point but rather
the instability of the solid phase which is prevented by
the melting process itself.\cite{bruesch}
In this perspective this criterion should not be employed
at a quantitative level. Nevertheless, since the solid phase instability
and the actual melting process are usually related to each other,
the PT theory provides a simple and useful way to get information
about the tendency of a system towards melting and surface melting
and their dependence on microscopic parameters.
%
% More detailed analyses are however required for a quantitative study.

In this paper we generalize the results of the PT approach in the
case of quantum solids. The Lindemann criterion in the quantum
solid is shown to be twice as large as the one in the classical limit,
in agreement with experimental reports.\cite{polturak}
We show a phase diagram for both the
bulk and surface melting cases and we investigate also the local melting
due to an isotopic substitution.
The temperature dependence of the lattice fluctuations for the
different classical/quantum regimes is evaluated and
also the profile of the lattice fluctuations as function
of the distance from the surface or the isotopic impurities.
The paper is organized as follows: in Sect. \ref{s-pt}, we review
the approach of Pietronero-Tosatti for classical solids; in
\ref{s-bulk} we generalize the PT approach to investigate
bulk properties in quantum solids; surface melting and solid phase
instability close to a quantum isotope impurity are analyzed
respectively in  Sect. \ref{s-sm} and Sect. \ref{s-qimp}.
Finally, in  Sect. \ref{s-concl}, we discuss our results
and draw some final conclusions.

\section{SCHA and solid phase instability in bulk and
on surfaces}
\label{s-pt}

Let us consider for simplicity a one-dimensional chain
of atoms. At the harmonic level, we can write the equations of motion for
the lattice displacement $u_n$:
\begin{eqnarray}
M \ddot{u}_n
+ 
\frac{k_{n,n+1}}{2} (u_n-u_{n+1}) + \frac{k_{n,n-1}}{2} (u_n-u_{n-1}) =0,
\label{osc}
\end{eqnarray}
where $M$ is the atomic mass,
$n$ denotes the site index. The constant forces $k_{n,n'}$,
at the harmonic level, are related to the inter-atom potential $V_{n,n'}$
through the relation
$k_{n,n'}=\partial^2 V_{n,n'}/\partial u_n^2|_{\{u_n\}=0}$.
Writing the potential $V_{n,n'}$ in terms of a Fourier expansion,
$V_{n,n'}=\sum_q V_q \exp[iq(u_n-u_{n'})]$, we have thus,
at the harmonic level,
$k_{n,n'}= k_0 = -\sum_q q^2 V_q$.

In the spirit of SCHA, anharmonic terms can be taken into account,
by replacing the constant forces $k_{n,n'}$, evaluated at
the lattice equilibrium, with their expectation value $\tilde{k}_{n,n'}$
averaged over the lattice fluctuations. We have thus explicitly:
\begin{eqnarray}
\tilde{k}_{n,n'}
&=&
\left\langle \frac{\partial^2 V_{n,n'}}{\partial u_n^2}\right\rangle
\nonumber\\
&=&
-\sum_q q^2 V_q \exp[-q^2\langle|u_n-u_{n'}|^2\rangle/2]
\nonumber\\
&\simeq&
k_0 \exp[-\lambda\langle u_n^2\rangle/2-\lambda\langle u_{n'}^2\rangle/2],
\label{kan}
\end{eqnarray}
where in the last line we have neglected the cross terms and
we have replaced the dependence on the momenta in the exponential
with an effective parameter $\lambda$.

By inserting (\ref{kan}) in Eq. (\ref{osc}) and
considering the motion of each atom as an Einstein oscillator
we have:
\begin{eqnarray}
M\ddot{u}_n
+ \frac{1}{2} \left[\tilde{k}_{n,n+1}+\tilde{k}_{n,n-1}\right]u_n=0,
\label{scha}
\end{eqnarray}
where anharmonic effects are taken into account in the self-consistent
renormalization of the elastic constants $\tilde{k}_{n,n'}$.
Note that $\tilde{k}_{n,n'}$ depends on the expectation
value of the quadratic lattice fluctuations on both sites $n$, $n'$.
It follows that the atomic motion described in Eq. (\ref{scha}) is ruled
by the lattice fluctuations of the lattice {\em environment}.
In a bulk system $\langle u_n^2\rangle=\langle u_{n'}^2\rangle
=\langle u^2\rangle$, then
\begin{equation}
\label{eqn:k-tilde}
\tilde{k}_{n,n'}
= 
\tilde{k} = k_0 \exp\left[-\lambda \langle u^2 \rangle\right].
\end{equation}
and we get an unique self-consistent equation
\begin{eqnarray}
\langle u^2\rangle
&=&
\frac{k_{\rm B}T}{\tilde{k}}
%\nonumber\\
%&=&
=
\frac{k_{\rm B}T}{k_0}
\exp\left[\lambda\langle u^2\rangle\right],
\label{self_bulk1}
\end{eqnarray}
where $k_{\rm B}$ is the Boltzmann constant.
In similar way, the SCHA phonon frequency is given by
$\tilde{\omega}_0=\sqrt{\tilde{k}/M}=
\omega_0\exp[-\lambda\langle u^2\rangle/2]$, where
$\omega_0=\sqrt{k_0/M}$ is the bare phonon frequency at
the purely harmonic level.
It is convenient to rewrite Eq. (\ref{self_bulk1}) by introducing
the dimensionless quantities $y=\lambda\langle u^2\rangle$,
$\tau_{\rm cl}=\lambda k_{\rm B}T/k_0$:
\begin{eqnarray}
y(\tau_{\rm cl})
&=&
\tau_{\rm cl}\mbox{e}^{y(\tau_{\rm cl})}.
\label{self_bulk}
\end{eqnarray}
Eq. (\ref{self_bulk}) has no solution for
$\tau_{\rm cl} \ge \tau_{\rm cl}^{\rm max} = 1/\mbox{e}=0.368$,
which determines a critical temperature $k_{\rm B} T_c=0.368 k_0/\lambda$.
At this value $y(\tau_{\rm cl}^{\rm max})=1$ and
the maximum magnitude of the allowed lattice fluctuations above which
the solid phase is unstable is
$\langle u^2 \rangle^{\rm max}=1/\lambda$.
Note that $\langle u^2 \rangle^{\rm max}$  depends neither
on the atomic mass nor on the force constant $k_0$, in agreement
with the observation of a material independent Lindemann criterion.

Eq. (\ref{scha}) represents also the starting point
to apply the SCHA to surface melting.
In this case,
one defines a local average lattice fluctuation
$\langle u_n^2 \rangle$
%($y_n$), 
which depends on the site index $n$.
In the same spirit one can define a local elastic constant:
%the effective elastic constant of the site $n$
%results to depend also on the local lattice fluctuations of the $n-1$ and
%$n+1$ atoms, namely:
\begin{eqnarray}
 \tilde{k}_{n,n-1,n+1}
\!&=&\!
 \left[ \tilde{k}_{n,n+1}+\tilde{k}_{n,n-1} \right]
\nonumber\\
\!&=&\!
k_0\mbox{e}^{-\lambda \langle u_n^2 \rangle/2}
\!\!\left[
\mbox{e}^{-\langle u_{n-1}^2 \rangle/2}
\!+\!
\mbox{e}^{-\langle u_{n+1}^2 \rangle/2}
\right] \label{k3}
\end{eqnarray}
We can write thus a set of recursive equations where
the lattice fluctuations of the atom $n$ depend
on the lattice fluctuations of the $n-1$ and $n+1$ atoms.
The recursion is truncated at the atom $n=1$ which represents
the outer atom close to the free surface. This atom probes an effective
harmonic potential smaller than the bulk, which increases its tendency
towards melting. A numerical solution shows that the solid phase
for the surface atoms becomes unstable at $\tau_{\rm cl}^{\rm SM}=0.271$,
26 \% smaller than the bulk value.
The same theory permits to evaluate the profile of the lattice fluctuations
as function of the distance from the surface.
These theoretical predictions agree quite well with the
profile of the lattice fluctuations close to defects (grain boundaries,
dislocations, vacancies) in colloidal solids.\cite{yodh}
Note that, although the temperature of surface melting is smaller
than in the bulk, local lattice fluctuations of the outer atoms
can be {\em larger} than the ones in the bulk, violating locally
the Lindemann criterion. This is also in agreement
with Ref. [\onlinecite{yodh}]. For instance, for the outer atoms $n=1$
one finds $y_1^{\rm SM}=1.74$. This is 74 \% larger than the value in the
bulk.

\section{Bulk properties of quantum solids}
\label{s-bulk}

We generalize now the above theory to the case
of quantum solids.
In the following we shall assume 
a one-particle picture to be still valid,
because of the smallness of the exchange
terms in the solid phase ($J^{\rm max} \sim  0.1$ K in $^4$He,
$J^{\rm max} \sim$ $\mu$K in $^3$He)
with respect to the melting temperatures
$T_{\rm m} \gtrsim 2$ K.\cite{ceperley_He3,ceperley_He4}
On the other hand a major role in our approach will be
played by the
quantum fluctuations which dominate at low temperature
in the quantum regime.
According to this perspective,
the atomic motion
of the atom $n$ is described in terms of the SCHA Hamiltonian
of the quantum oscillator:
\begin{eqnarray}
\left[-\frac{\hbar^2\nabla^2_u}{2M}
+\frac{1}{4}
\tilde{k}_{n,n-1,n+1} u^2_n \right] \Psi \left( u_n \right)
& = & E \Psi \left( u_n \right),
\label{qscha}
\end{eqnarray}
where the self-consistent expression
for the local potential $\tilde{k}_{n,n-1,n+1}$
is reported in Eq. (\ref{k3}).

We consider first the melting properties of bulk systems 
($\tilde{k}_{n,n-1,n+1} = 2  \tilde{k}$).
In this SCHA quantum model the total amount
of lattice fluctuations is now easily computed as:
\begin{equation}
\langle u^2 \rangle
=
\frac{\hbar}{2 M \tilde{\omega}_0} 
\left[1+2n\left(\frac{\hbar \tilde{\omega}_0}{k_{\rm B}T} \right)\right],
\label{gen}
\end{equation}
where $n(x)=1/[\mbox{e}^x-1]$ is the Bose factor and where
we remind $\tilde{\omega}_0 = \sqrt{\tilde{k}/M}$ and 
$\tilde{k}$ is given by Eq.\ref{eqn:k-tilde}. 
In the classic limit $k_{\rm B} T \gg \hbar \tilde{\omega}_0$,
$n(x)\simeq 1/x\gg 1$, and we recover the classical result
of Eq. (\ref{self_bulk1}).
On the other hand, in the zero temperature limit, lattice fluctuations
are due only to zero point quantum motion.
In this case $n(x)=0$ and Eq.(\ref{gen}) reads:
\begin{eqnarray}
\label{eq:uB-T0}
\langle u^2 \rangle
&=&
\frac{\hbar}{2 \sqrt{M \tilde{k} }} 
=
\frac{\hbar}{2 \sqrt{M k_0 }}
\exp\left[\lambda \langle u^2 \rangle/2\right],
\end{eqnarray}
which, introducing the variable
$\tau_{\rm Q}= \lambda \hbar/ 2\sqrt{k_0M}$,
can be written in the dimensionless form:
\begin{equation}
y(\tau_{\rm Q})=\tau_{\rm Q} \mbox{e}^{y(\tau_{\rm Q})/2}.
\label{qua}
\end{equation}

Eq. (\ref{qua}) represents the quantum generalization of Eq. (\ref{self_bulk})
where the instability of the solid phase is now triggered by the magnitude
of the quantum lattice fluctuations. This occurs for
$\tau_{\rm Q} \ge \tau_{\rm Q}^{\rm max} = 2/\mbox{e}=0.736$.
It is interesting to note that the breakdown of the solid phase
driven by quantum fluctuations is not merely equivalent
to the one related to the thermal motion. Indeed, for a quantum solid,
we would predict a maximum magnitude of lattice fluctuations
$y(\tau_{\rm Q}^{\rm max})=2$, {\em two times} larger than 
for classical solids.
This behavior is indeed in agreement with the report of
the Lindemann ratio $u_{\rm rms}/a \simeq 0.28$  
in helium solids \cite{polturak,wilks,glyde} 
to compare with $u_{\rm rms}/a \simeq 0.16$ for classical 
solids.
% (remind that $u_{\rm rms} \sim \sqrt{y^{\rm max}}$).

We also consider now the general case where both thermal and quantum
fluctuations are important.
From Eq. (\ref{gen}),
after few straightforward passages, we get
\begin{equation}
y(\tau_{\rm Q},\tau_{\rm cl})=
\tau_{\rm Q} \mbox{e}^{y(\tau_{\rm Q},\tau_{\rm cl})/2}
\left[1+2n\left(\frac{2\tau_{\rm Q}}{\tau_{\rm cl}}
\mbox{e}^{-y(\tau_{\rm Q},\tau_{\rm cl})/2}\right)
\right]. 
\label{tot}
\end{equation}

Eq. (\ref{tot}) generalizes the stability criterion based on the SCHA
in the full quantum-thermal case.
As a general rule we can expect that the classical regime is relevant
in the empirical range $k_{\rm B}T/\hbar \omega_0 \gtrsim 1/4$,
which corresponds to $\tau_{\rm Q} \lesssim 2 \tau_{\rm cl}$,
while in the opposite regime $\tau_{\rm Q} \gtrsim 2 \tau_{\rm cl}$
quantum effects are dominant.

\begin{figure}
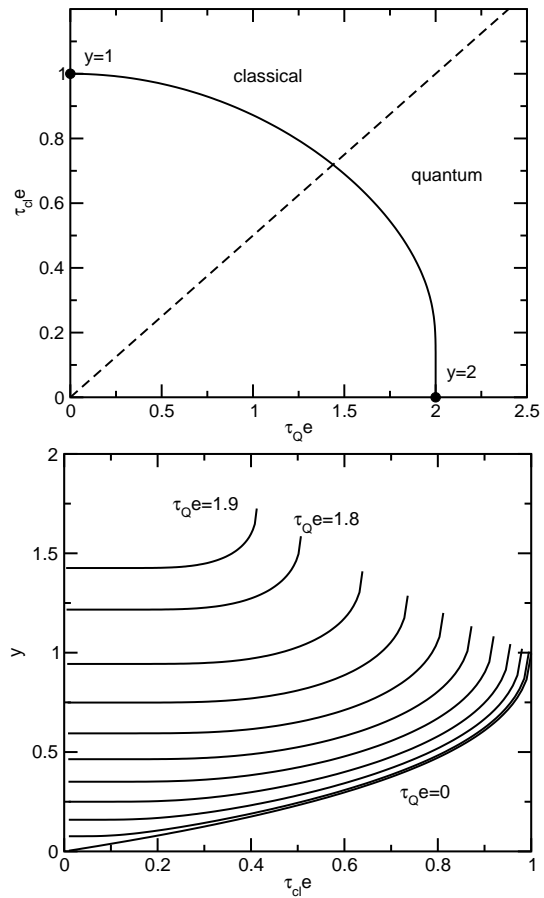

\centerline{\psfig{figure=f-bound2.eps,width=7cm,clip=}}
%\hspace{4mm}
\centerline{\psfig{figure=f-u2_cl.eps,width=7cm,clip=}
}
\caption{(top panel)
Phase boundary of the SCHA in the $\tau_{\rm Q}$-$\tau_{\rm cl}$ space;
(bottom panel) Lattice fluctuations $y=\lambda \langle u^2 \rangle$
as function of the classical parameter $\tau_{\rm cl}$ for
(from the bottom to the top)
$\tau_{\rm Q}\mbox{e}=0.2, 0.4, 0.6, \ldots, 1.6, 1.8, 1.9$
(we remind that $\tau_{\rm Q}^{\rm max}=2/\mbox{e}$ ).}
\label{f-bound}
\end{figure}
In Fig. \ref{f-bound}, we show the phase diagram in the
full $\tau_{\rm Q}$-$\tau_{\rm cl}$ space where the instability of the SCHA
occurs. Along the boundary line,
the critical lattice fluctuations increase smoothly from $y=1$
in the $\tau_{\rm Q}=0$ case to $y=2$
in the $\tau_{\rm cl}=0$ case. Also interesting is the dependence
of the lattice fluctuations
as function of $\tau_{\rm cl}$, namely the temperature 
(Fig. \ref{f-bound}, bottom panel).
In the classical case, $\tau_{\rm Q}=0$, 
the quadratic fluctuations $y \propto \langle u^2 \rangle$
increase linearly with $\tau_{\rm cl}$ until anharmonic effects take place.
Anharmonicity is reflected in a upturn of the temperature
dependence of $y(\tau_{\rm cl})$ and eventually in the breakdown
of the solid phase for $\tau_{\rm cl}=1/\mbox{e}$ and $y=1$.
Increasing $\tau_{\rm Q}$ leads not only to the presence
of zero point motion quantum fluctuations at $\tau_{\rm cl}=0$,
but also to an overall change of the temperature dependence of $y$.
In particular, the range of the linear temperature dependence,
characteristic of classical harmonic solids, is rapidly reduced and
for strongly quantum solids it disappears. Lattice fluctuations
are large {\em already} at $T=0$ and they
are almost constants in a wide temperature range (note that in this
regime anharmonic effects are in any case present
due to quantum fluctuations)
until an abrupt upturn with the temperature leads to the
breakdown of the solid phase.
This trends is in good qualitative agreement
with recent experimental measurements\cite{arms}
and Quantum Monte Carlo calculations.\cite{draeger}
We shall discuss them in details in Sect. \ref{s-concl}.

\section{Surface melting of quantum solids}
\label{s-sm}

After having investigated the bulk properties of quantum solids,
we analyze now the occurrence the role of quantum fluctuations
on the surface melting.

We can write a recursive set of equations
by considering the quantum/thermal SCHA solution
of the $n$-th atom
\begin{equation}
\langle u^2_n \rangle
=\frac{\hbar}{2 M \tilde{\omega}_n }
\left[1+2n\left(\frac{\hbar \tilde{\omega}_n}{k_{\rm B}T}\right)\right],
\label{genq}
\end{equation}
where $\tilde{\omega}_n=\sqrt{\tilde{k}_{n,n-1,n+1}/2 M}$
and where the local elastic constant $\tilde{k}_{n,n-1,n+1}$
is still given by Eq. (\ref{k3}).
% which is valid as well in the
%generic quantum case.
Employing the usual dimensionless variables
$\tau_{\rm Q}$, $\tau_{\rm cl}$, $y_n$, we can thus write:
\begin{eqnarray}
y_n
&=&
\frac{\sqrt{2}\tau_{\rm Q}\mbox{e}^{y_n/4}}
{\sqrt{\mbox{e}^{-y_{n-1}/2}+\mbox{e}^{-y_{n+1}/2}}}
\nonumber\\
&&\times
\left[
1+2n\left(
\frac{2\tau_{\rm Q}}{\tau_{\rm cl}}
\frac{\sqrt{\mbox{e}^{-y_{n-1}/2}+\mbox{e}^{-y_{n+1}/2}}}
{\sqrt{2}\mbox{e}^{y_n/4}}
\right)
\right],
\label{ynsm}
\end{eqnarray}
which is valid for any $n \ge 2$, while the outer atom $n=1$
obeys the relation
\begin{eqnarray}
y_1
\!&=&\!
\sqrt{2}\tau_{\rm Q}\mbox{e}^{(y_1+y_2)/4}
\!\left[
1\!+\!2n\!\left(
\frac{2\tau_{\rm Q}}{\sqrt{2}\tau_{\rm cl}}\mbox{e}^{-(y_1+y_2)/4}
\right)
\!\right].
\label{y1sm}
\end{eqnarray}

In order to obtain a numerical solution of Eqs. (\ref{ynsm})-(\ref{y1sm})
for given $\tau_{\rm Q}$, $\tau_{\rm cl}$
in the stable solid phase,
we start by choosing a trial
value of $y_1$. The full set of $\{y_n\}$ is thus obtained
by  Eqs. (\ref{ynsm})-(\ref{y1sm}). The initial trial value
of $y_1$ is thus varied until $y_{n=\infty}$ converges
to its bulk value. Typically, this is  the only physical solution,
since $y_{n=\infty}$ diverges for larger values of $y_1$
while it becomes rapidly negative for smaller values of $y_1$.
For $\tau_{\rm Q}$, $\tau_{\rm cl}$ larger than some critical value,
the procedure does not converge for {\em any} value
of $y_1$, signalizing that the solid phase of the surface atom,
described by the SCHA, is unstable.

\begin{figure}
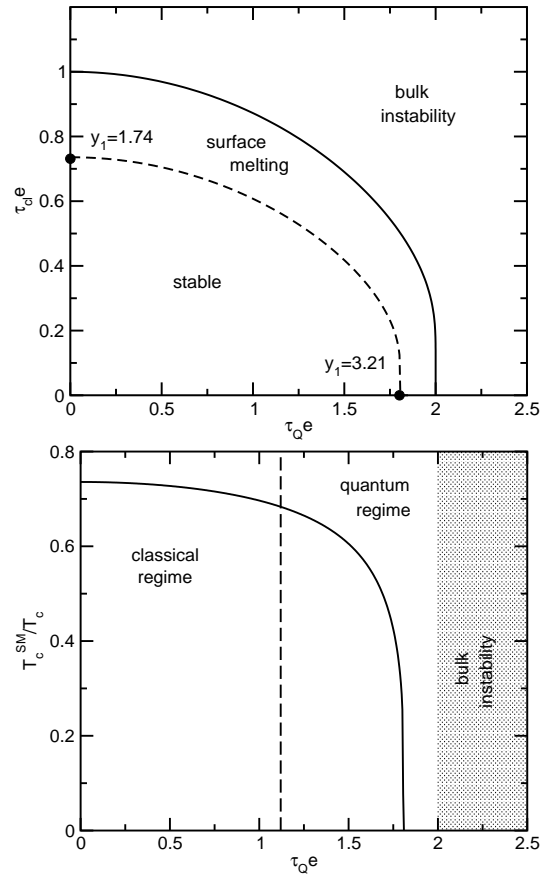

\centerline{\psfig{figure=f-bound_sm2.eps,width=7cm,clip=}}
%\hspace{4mm}
\centerline{\psfig{figure=f-ratio_sm.eps,width=7cm,clip=}}
\caption{(top panel)
Phase boundary for the surface melting instability (dashed line)
compared with the bulk instability (solid line)
in the $\tau_{\rm Q}$-$\tau_{\rm cl}$ space;
(bottom panel) Ratio between surface melting temperature
$T_c^{\rm SM}$ and bulk melting temperature $T_c$ as function
of $\tau_{\rm Q}$. For $\tau_{\rm Q}\mbox{e}$
even the bulk phase is unstable. For 
$1.12 < \tau_{\rm Q}\mbox{e} < 2$ the system is in a quantum regime
where $k_{\rm B}T_c^{\rm SM}/\hbar \tilde{\omega}_0 \lesssim 1/4$.}
\label{f-sm_phd}
\end{figure}

The resulting phase diagram, in the 
full $\tau_{\rm Q}$-$\tau_{\rm cl}$ space,
is shown in Fig. \ref{f-sm_phd} (top panel), where we compare the boundary
of the surface melting instability (dashed line)
with the one of the bulk melting (solid line).
For the pure quantum case, $\tau_{\rm cl}=0$, at zero temperature
the surface instability occurs for  $\tau_{\rm Q}^{\rm SM}=0.664$ where
the lattice fluctuations of the outer atoms become as large as
$y_{1,\rm Q}^{\rm SM}=3.21$.
It is interesting to notice that, for $0.664 < \tau_{\rm Q} < 0.736$,
the surface is unstable {\em even} at zero temperature
whereas the bulk solid phase is always stable up to
a finite temperature range.
The ratio $T_c^{\rm SM}/T_c$
between the surface melting temperature
and the temperature of bulk melting is shown in the bottom panel
of FIG. \ref{f-sm_phd} showing that the critical temperature
of surface melting can be significantly lower than the bulk one
in quantum solids.

Before concluding this section, we would like to
briefly compare the melting occurring at a free surface with
other cases such as grain boundaries.
In the case of a free surface, in going from Eq. (\ref{ynsm})
to Eq. (\ref{y1sm}),
we have dropped in Eq. (\ref{y1sm}) the contribution of the $n=0$ atom.
We note that the same results would be obtained in Eq. (\ref{ynsm})
considering $n=1$ and
assuming the lattice fluctuations at the site $n=0$ to be infinite,
namely $y_{n=0}=\infty$. This latter condition would be obtained by
the harmonic oscillator solution of (\ref{genq}) at the site $n=0$
with a vanishing elastic constant $\tilde{k}_{n,n-1,n+1}$, and
it express nothing more than the condition that atoms for $n < 1$
are not in a solid arrangement but in a gaseous phase.

An intermediate situation is encountered when melting at grain boundary
interfaces is considered.
In this case the outer atom $n=1$ of a grain would not probe
a free surface at the site $n=0$, but it will interact with
a lattice environment with a different arrangement.
These two situations can be described by a similar set of recursion 
relations (\ref{ynsm}) but with different boundary conditions:
in the free surface case boundary conditions at site $n=0$ will
be described by a completely soft oscillator $\tilde{k}_{n,n-1,n+1}=0$,
signalizing that bulk solid is interfaced with a free gaseous phase;
on the other hand, in the case of grain boundaries, the
outer atom $n=1$ will still probe a crystal structure for $n\le 0$,
although with a different arrangement. The boundary conditions at site $n=0$
will be still described thus by Eq. (\ref{genq}), but with a not completely
soft mode. We expect thus that melting processes occur as well
at grain boundaries as in the case of free surface.
From the mathematical point of view, this situation is identical
to the case of quantum isotopic substitutions, and it will be discussed
in details in the next section.

\section{Quantum melting driven by isotopic impurities}
\label{s-qimp}

In this section we address the problem of the solid phase stability
close to a single local
isotopic substitution embedded in a perfect lattice structure.
In the SCHA approach, local stability of the solid phase is given
by the solution of Eq. (\ref{genq}).
It is easy to check that, in the classical limit
$k_{\rm B}T \gg \hbar \tilde{\omega}_n$, the dependence on the atomic mass
$M$ in Eq. (\ref{genq}) drops out, so that different isotope solids should
probe the same stability conditions.
On the other hand,
the mere observation of a different melting line for $^4$He and $^3$He
is a direct evidence that helium is in a quantum
regime.\cite{hansen,boninsegni}
Different isotopes are thus expected to affect
the bulk solid phase stability.
We expect the same at the local level.

In the following we shall consider the case of a isolate
substitution with a lighter isotope in a host matrix of heavier atoms.
Quantum fluctuations in the two cases will be ruled locally by
the parameters $\tau_{\rm L}= \lambda \hbar/ 2 \sqrt{k_0M_{\rm L}}$
$\tau_{\rm H}= \lambda \hbar/ 2 \sqrt{k_0M_{\rm H}}$,
respectively for the lighter (L) and for the heavier (H) atoms.
To study the stability of the solid phase close
to this isotopic quantum impurity, we can still employ the
recursive relations (\ref{ynsm}), namely
for $n\le -1$, $n\ge 1$ we set $\tau_{\rm Q}=\tau_{\rm H}$,
whereas for $n=0$ (quantum isotope impurity) we have
$\tau_{\rm Q}=\tau_{\rm L}$.
We shall consider the representative case of a $^3$He impurity
embedded in $^4$He solid. In this case $\tau_{\rm L}/\tau_{\rm H}=\sqrt{4/3}$.

\begin{figure}
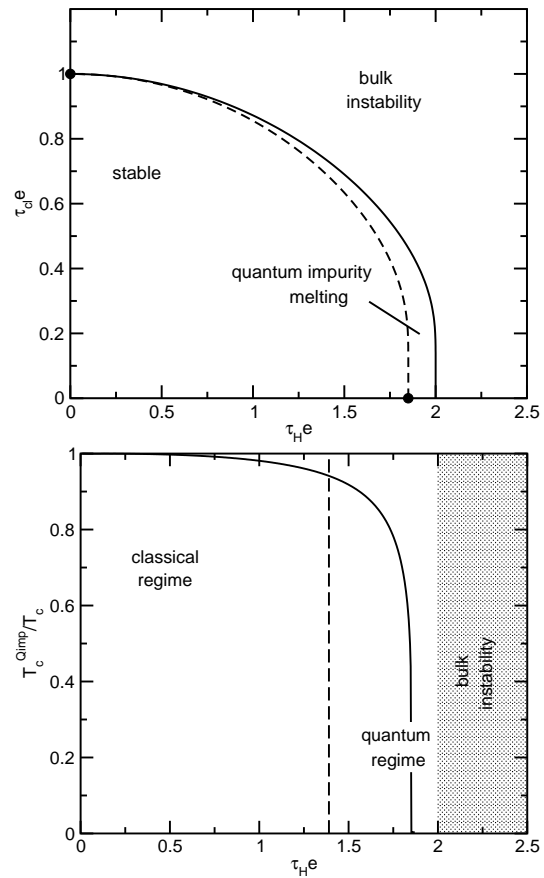

\centerline{\psfig{figure=f-bound_qimp2.eps,width=7cm,clip=}}
%\hspace{4mm}
\centerline{\psfig{figure=f-ratio_qimp.eps,width=7cm,clip=}}
\caption{(top panel)
Phase boundary for the lattice instability around
a quantum isotopic substitution with
$\tau_{\rm L}/\tau_{\rm H}=\sqrt{4/3}$ (dashed line)
compared with the bulk instability (solid line);
(bottom panel) Ratio between melting lattice temperature $T_c^{\rm Qimp}$
around the quantum impurity
and bulk melting temperature $T_c$ as function
of host quantum parameter $\tau_{\rm H}$.
In the quantum regime $1.39 < \tau_{\rm H}\mbox{e} < 2$,
where $k_{\rm B}T_c^{\rm Qimp}/\hbar \tilde{\omega}_0 \lesssim 1/4$,
the local melting temperature around the quantum
impurity is sensible lower than in the bulk, and
for $1.85 < \tau_{\rm H}\mbox{e} < 2$ solid phase
around isotopic quantum impurities is unstable even
at $T=0$.}
\label{f-qimp_phd}
\end{figure}

In Fig. \ref{f-qimp_phd}(top panel) we show the phase diagram
of the lattice instability of the {\rm host} $^4$He solid
close to the quantum isotopic $^3$He impurity.
It is instructive to compare the classical limit
$\tau_{\rm Q}=0$ with the pure quantum one
$\tau_{\rm cl}=0$. In the first case lattice fluctuations
of the guest atom, as well as of the host atoms,
are independent on the relative atomic mass and they depend
only on the temperature. As a consequence, the solid phase close
to the guest atom is completely unaffected
by the isotopic substitution.
A quite different situation occurs
in the highly quantum regime $\tau_{\rm cl}=0$.
In this case local quantum lattice fluctuations of the lighter guest
atom can be significantly enhanced due to its reduced atomic mass,
and they can be sufficiently large to induce a {\em local}
melting of the {\em host} solid phase.
At $\tau_{\rm cl}=0$ this occurs for  $\tau_{\rm H} > 0.681$,
not much higher than in the case of a free surface truncation
($\tau_{\rm Q} > 0.664$). Note that
Fig. \ref{f-qimp_phd} defines a region (quantum impurity melting)
where solid phase is still stable in the bulk but local
quantum lattice fluctuations break down the solid phase close
to the isotopic substitution.
On the physical ground we can expect liquid bubbles
of {\em host} atoms to appear close to the guest isotope.
Unfortunately, since the present analysis is only related to the
stability condition of the solid phase, we are not able to estimate
the size of the liquid bubble, and more sophisticated approaches
are needed.
It is interesting to note that, for quantum solids,
the critical temperature $T_c^{\rm Qimp}$
for the local stability of the solid phase close to the
quantum isotope impurity is reduced with respect to
the bulk $T_c$. This is shown in the bottom panel of Fig. \ref{f-qimp_phd}
where the ratio between the local $T_c$ close to the impurity and
the bulk $T_c$ is plotted as function of the quantum degree
of the system, parametrized by $\tau_{\rm H}$. In the quantum regime,
where $T_c^{\rm Qimp} \lesssim \hbar \tilde{\omega}_0/4$, the local melting
temperature $T_c^{\rm Qimp}$ 
can be significantly lower than the one in the bulk $T_c$,
and, for $1.39 < \tau_{\rm H}\mbox{e} < 2$, we expect a quantum isotopic
impurity to induce local melting down to $T=0$, although
the bulk phase is still stable.

\section{Discussion and Conclusions}
\label{s-concl}

In this paper we have investigated the stability of quantum solids
with respect to surface melting and to isotopic quantum substitutions.
Both these phenomena can be essentially related to the
amount of lattice fluctuations, and they can be driven thus by thermal
fluctuations as well as by the zero point quantum motion.
We have shown that the effects of isotopic impurities
and surface melting are strongly enhanced in quantum solids.
In particular we show that
when quantum fluctuations are dominant
in quantum solids the solid phase can be
rapidly destroyed on the surface and close to quantum impurities
at temperatures much smaller than for the bulk melting.

Helium solids are the natural candidates
where the quantum instabilities of surface or 
interface can occur.
The actual relevance of these quantum melting effects
are of course ruled by the magnitude of the
quantum lattice fluctuations which are parametrized
in our model by the quantity $\tau_{\rm Q}$.
An accurate calculation of the quantum lattice fluctuations
as a function of the temperature in $^4$He and $^3$He solids
has been provided recently,
by using of Quantum Monte Carlo (QMC) techniques,
by Draeger and Ceperley in Ref. \onlinecite{draeger},
in excellent agreement with the experimental data.\cite{arms} 
Quite interestingly, they find that the mean square lattice displacement
$\langle u^2\rangle_T$
does not follow at low temperature an harmonic behavior
$\langle u^2\rangle_T \simeq \langle u^2\rangle_{T=0}+\alpha T^2$,
but rather a more shallow one 
$\langle u^2\rangle_T \simeq \langle u^2\rangle_{T=0}+\beta T^3$.

\begin{figure}[t]
\centerline{\psfig{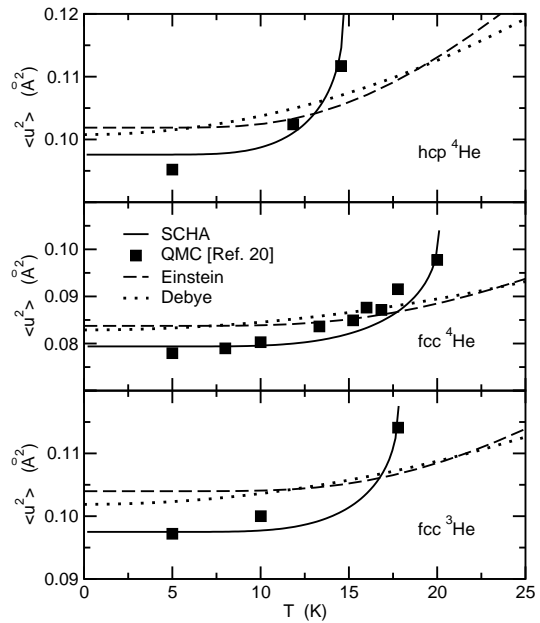}}
%\centerline{\psfig{figure=art_fit_He4_FCC.eps,width=7cm,angle=270.,clip=}}
%  \centerline{\psfig{figure=art_fit_He3_FCC.eps,width=7cm,angle=270.,clip=}}
% (bottom panel)  $^3$He FCC phase at molar volume $12.12 cm^3/mole$
\caption{Lattice fluctuations $\langle u^2\rangle$
evaluated within the SCHA (solid lines)
as function of temperature for different helium solid conditions
compared with Quantum Monte Carlo data of Ref. \onlinecite{draeger}.
Values of $k_0$ and $\lambda$ in SCHA obtained by fitting QMC data
are reported in Table \ref{table}.
Also shown are the purely harmonic fitting of the QMC data with a
Einstein and a Debye model.}
\label{u2_helium4}
\end{figure}

Ref. \onlinecite{draeger} represents a suitable
source to estimate an effective value of $\tau_{\rm Q}$
representative of solid helium.
To this aim we fit the temperature dependence
of the QMC data of
Ref. \onlinecite{draeger} with our quantum SCHA model 
described by Eq. (\ref{tot}), where only two independent
fitting parameters appear, namely $\lambda$ and $k_0$
(remind that $\tau_{\rm cl}=\lambda k_{\rm B}T/k_0$,
$\tau_{\rm Q}= \lambda \hbar/ 2\sqrt{k_0M}$).
The fit of our quantum SCHA [Eq. (\ref{tot})]
compared with the QMC data is shown in Fig. \ref{u2_helium4}
for three representative cases where
the number of numerical data is larger than the number of independent
fitting parameters to guarantee the significance of the fitting
procedure. Also shown is the fit with a purely harmonic
model obtained by setting $\lambda=0$.
The extracted values of $\lambda$ and $k_0$, as well
as of the corresponding $\tau_{\rm Q}$ and of the anharmonic
renormalized phonon frequency at $T=0$ $\tilde{\omega}_0$
are reported in Table \ref{table}, where also
we report
the critical temperature $T_c$ for the solid phase
bulk instability evaluated within the SCHA and the experimental
melting temperature $T_{\rm m}^{\rm exp}$.\cite{wilks,dobbs}
\begin{table}[b]
\begin{tabular}{|c|c|c|c|}
\hline
\hline
                         & hcp $^4$He      & fcc $^4$He      & fcc $^3$He      \\
\hline
$V_0$ (cm$^3$/mole)      & 12.12           & 10.98           & 11.54           \\
\hline
$k_0$ (meV/\AA$^2$)      & $110 \pm 10$    & $140 \pm 10$   & $150 \pm 10$     \\
\hline
$\lambda$ (\AA$^{-2}$)   & $14 \pm 1$      & $15.2 \pm 0.8$  & $14.7 \pm 0.7$  \\
\hline
$\tau_{\rm Q}$           & $0.69 \pm 0.08$ & $0.66 \pm 0.06$ & $0.70 \pm 0.06$ \\
\hline
$\tilde{\omega}_0$ (meV) & $4.6 \pm 0.1$   & $5.4 \pm 0.4$   & $6.1 \pm 0.5$   \\
\hline
$T_c$ (K)                & $14 \pm 4$      & $20 \pm 4$      & $18 \pm 4$      \\
\hline
$T_m^{\rm exp}$ (K)      & $\sim 15$      & $\sim 21$       & $\sim 22$        \\
\hline
\hline
\end{tabular}
\caption{Values of $k_0$ and $\lambda$
in SCHA obtained by fitting the QMC data of 
Ref. \onlinecite{draeger} for three representative helium solids,
namely: hcp $^4$He at molar volume $V_0=12.12$ cm$^3$/mole,
fcc $^4$He at molar volume $V_0=10.98$ cm$^3$/mole,
and 
fcc $^3$He at molar volume $V_0=11.54$ cm$^3$/mole.
Also reported are the corresponding values of $\tau_{\rm Q}$,
the renormalized phonon frequency $\tilde{\omega}_0$ and the predicted
critical temperature $T_c$ of the solid phase bulk instability
compared with the experimental melting temperature
$T_{\rm m}^{\rm exp}$.\cite{wilks,dobbs}}
\label{table}
\end{table}

It is worth to comment about the
temperature behavior of the QMC data
compared with the harmonic ($\lambda=0$)
and anharmonic SCHA fit.
An important point to be here underlined is that
QMC results show a {\em large} mean square lattice displacement
at zero temperature all together with a rapidly turn up
of $\langle u^2 \rangle$ close to the solid bulk instability.
As we have discussed in Sect. \ref{s-bulk}, this is
a characteristic trend of highly quantum solids.
On the other hand, this behavior is poorly reproduced
by a purely harmonic model where the amount of the
lattice fluctuations at $T=0$ is inversely proportional to the
temperature dependence. This is even more true
if a Debye model would be employed since the temperature
dependence of a Debye model is even more shallow 
than in the Einstein case.

The strong quantum degree of solid helium,
qualitatively predicted by these arguments,
is confirmed by the numerical analysis
of the SCHA fit which predicts a quantum parameter
$\tau_{\rm Q}$ in the range $\tau_{\rm Q} \simeq 0.66-0.7$
for the three samples here considered.
The robustness of our fits is confirmed by the nice agreement
between the critical temperature for the bulk instability
of the solid phase estimated by the SCHA
and the experimental melting temperature.

These results have important consequences with respect
to the surface/grain-boundary melting instability and
local melting induced by quantum isotopic impurities.
The values of $\tau_{\rm Q}\simeq 0.69$, for the low pressure/
high molar volume $V_0=12.12$ cm$^3$/mole,
is safely larger than the value
$\tau_{\rm Q}^{\rm SM}\simeq 0.664$ where
surface melting occurs at zero temperature,
and also or the same order and slightly larger even
than $\tau_{\rm Q}^{\rm SM}\simeq 0.681$ where
isotopic impurity induced melting also
occurs at zero temperature.
Although these estimates have to be meant only indicative
of the quantum degree of helium solid, they clearly point out
that quantum anharmonic effects are large enough in solid helium,
for these or larger molar volumes, to enforce surface melting and local
melting close to quantum impurities {\em down to zero temperature}.
Quantum Monte Carlo simulations have actually confirmed premelting 
at surface between helium solid and Vycor walls \cite{khairallah} 
and internal interfaces of a pure helium system \cite{pollet}, 
although not all possible interfaces undergoes a solid/liquid transition.

These results shed an interesting light
also on the recent report of
the Non-Classical Rotational Inertia (NCRI)
observed in $^4$He.\cite{kc1,kc2}
While it was initially claimed to be an evidence of
a supersolid (SS) phase, 
subsequent experiments showed a strong dependence
of the NCRI on the annealing process,\cite{rittner}
on the presence of grain boundaries,\cite{sasaki}
on the amount of $^3$He concentration\cite{kc2,kc3,newchan}
as well as on the freezing procedure.\cite{kc3,newchan}
These observations give rise to
an alternative hypothesis to the SS phase, namely, that a liquid phase
is confined at the grain boundaries and that mass flow is related
to superfluidity of the liquid
component.\cite{burovski}
Our results confirm this scenario and
shed new perspectives about the
role of disorder/grain boundaries in solid helium.
In particular we provide a natural explanation
for the existence of a liquid (and thus probably
superfluid) phase at the grain boundaries
and we {\em predict} a local liquid phase
also around $^3$He impurities.
Local melting close to isotopic $^3$He impurities
should be thus explicitly considered.

\acknowledgments

This work was supported by the
Italian Research Programs MIUR PRIN 2005 and PRIN-2007.
G.R. acknowledges useful discussions with 
M. Holzmann and  D.M. Ceperley.


\begin{thebibliography}{99} 

\bibitem{dash}
J.G. Dash, Rev. Mod. Phys., {\bf 71}, 1737 (1999).

\bibitem{lowen}  
H. Lowen, Phys. Rep., {\bf 237}, 249 (1994).

\bibitem{forsblom}
M. Forsblom and G. Grimvall, Nature Materials {\bf 4}, 388 (2005).

\bibitem{lindemann}
F. Lindemann, Z. Phys. {\bf 11}, 609 (1910).

\bibitem{gilvarry}
J.J. Gilvarry, Phys. Rev. {\bf 102}, 308 (1956).

\bibitem{ross}
M. Ross, Phys. Rev. {\bf 184}, 233 (1969).

\bibitem{grimvall_ps}
G. Grimvall and S. Sj\"odin,
Phys. Scr. {\bf 10}, 340 (1974).

\bibitem{jin}
Z.H. Jin, P. Gumbsch, K. Lu, and E. Ma,
Phys. Rev. Lett. {\bf 87}, 055703 (2001).

\bibitem{yodh}
A.M. Alsayed, M.F. Islam, J. Zhang, P.J. Collings, and A.G. Yodh,
Science {\bf 309}, 1207 (2005).

\bibitem{pt}
L. Pietronero and E. Tosatti,
Solid State Commun. {\bf 32}, 255 (1979).

\bibitem{trayanov}
A. Trayanov and E. Tosatti,
Phys. Rev. B {\bf 38}, 6961 (1988).

\bibitem{tosatti-review}
For a review see:
U. Tartaglino, T. Zykova-Timan, F. Ercolessi, and E. Tosatti,
Phys. Rep. {\bf 411}, 291 (2005).

\bibitem{bruesch}
P. Br\"uesch,
{\em Phonons: Theory and Experiments III},
Springer Series in Solid State Sciences, v. {\bf 66}
(Springer, Berlin, 1987).

\bibitem{polturak}
E. Polturak and N. Gov,
Contem. Phys. {\bf 44}, 145 (2003).

\bibitem{ceperley_He3}
D.M. Ceperley and  G. Jacucci,
Phys. Rev. Lett. {\bf 58}, 1648 (1987).

\bibitem{ceperley_He4}
D.M. Ceperley and  B. Bernu,
Phys. Rev. Lett. {\bf 93}, 155303 (2004).

\bibitem{wilks}
J. Wilks,
{\em The Properties of Liquid and Solid Helium},
(Claredon Press, Oxford, 1967).

\bibitem{glyde}
H.R. Glyde,
Encyclopedia Phys. {\bf 1}, 1001 (2005).

\bibitem{arms}
D.A. Arms, R.S. Shah, and R.O. Simmons,
Phys. Rev. B {\bf 67}, 094303 (2003).

\bibitem{draeger}
E.W. Draeger and D.M. Ceperley,
Phys. Rev. B {\bf 61}, 12094 (2000).

\bibitem{hansen}
P. Loubeyre and J.P. Hansen 
Phys. Lett. A {\bf 80}, 181 (1980).

\bibitem{boninsegni}
M. Boninsegni, C. Pierleoni, and D. M. Ceperley,
Phys. Rev. Lett. {\bf 72}, 1854 (1994).

\bibitem{dobbs}
E.R. Dobbs,
{\em Helium Three},
(Oxford Press, Oxford, 2000).

\bibitem{khairallah}
S.A. Khairallah and D.M. Ceperley,
Phys. Rev. Lett. {\bf 95}, 185301 (2005).

\bibitem{pollet}
L. Pollet, M. Boninsegni, A.B. Kuklov, N.V. Prokof'ev,
B.V. Svistunov, and M. Troyer,
Phys. Rev. Lett. {\bf 98}, 135301 (2007). 

\bibitem{kc1}
E. Kim and M.H.W. Chan,
Nature {\bf 427}, 225 (2004).

\bibitem{kc2}
E. Kim and M.H.W. Chan,
Science {\bf 305}, 1941 (2004).

\bibitem{rittner}
A.S.C. Rittner and J.D. Reppy,
Phys. Rev. Lett. {\bf 98}, 175302 (2007).

\bibitem{sasaki}
S. Sasaki {\em et al.},
Science {\bf 313}, 1098 (2006).

\bibitem{kc3}
A.C. Clark, J.T. West, and M.H.W. Chan,
arXiv:0706.0906v2 [cond-mat.other] (2007).

\bibitem{newchan}
E. Kim, J.S. Xia, J.T. West, X. Lin, A.C. Clark, and M.H.W. Chan,
arXiv:0710.3370[cond-mat.other] (2007).

\bibitem{burovski}
E. Burovski, E. Kozik, A. Kuklov, N. Prokof'ev, and B. Svistunov,
Phys. Rev. Lett. {\bf 94} 165301 (2005). 

%  
%  \bibitem{prokofev1}
%  N.  Prokof'ev and B. Svistunov,
%  Phys. Rev. Lett. {\bf 94}, 155302 (2005).
%  
%  \bibitem{prokofev2}
%  N.  Prokof'ev,
%  Adv. Phys. {\bf 56}, 381 (2007).
%  



\end{thebibliography}
\end{document}